\documentclass[twocolumn,superscriptaddress,showpacs,preprintnumbers,amsmath,amssymb,nofootinbib]{revtex4-1}

\usepackage{latexsym}
\usepackage{graphicx}
\usepackage{dcolumn}

\newcommand{\bra}[1]{\langle #1 |}

\newcommand{\ket}[1]{| #1 \rangle}

\newcommand{\braket}[2]{\langle #1 | #2 \rangle}

\newcommand{\im}{\dot{\iota}\,}
\newcommand{\ketv}[1]{\ket{\mathbf{V}_{#1}}}
\newcommand{\brav}[1]{\bra{\mathbf{V}_{#1}}}
\newcommand{\ketvt}[1]{\ket{\widetilde{\mathbf{V}}_{#1}}}
\newcommand{\bravt}[1]{\bra{\widetilde{\mathbf{V}}_{#1}}}
\newcommand{\braketv}[2]{\langle \mathbf{V}_{#1} | \mathbf{V}_{#2} \rangle}
\begin{document}

\title{Dynamical entanglement purification using chains of atoms
       and optical cavities}

\author{Denis Gon\c{t}a}
\email{denis.gonta@mpl.mpg.de}

\author{Peter van Loock}
\email{peter.vanloock@mpl.mpg.de}

\affiliation{Optical Quantum Information Theory Group,
             Max Planck Institute for the Science of Light,
             G\"{u}nther-Scharowsky-Str. 1, Building 26,
             91058 Erlangen, Germany}

\date{\today}

\begin{abstract}
In the framework of cavity QED, we propose a practical scheme to purify 
dynamically a bipartite entangled state using short chains of atoms 
coupled to high-finesse optical cavities. In contrast to conventional
entanglement purification protocols, we avoid CNOT gates, thus reducing 
complicated pulse sequences and superfluous qubit operations.
Our interaction scheme works in a deterministic way, and together
with entanglement distribution and swapping, opens a route towards 
efficient quantum repeaters for long-distance quantum 
communication.
\end{abstract}


\maketitle

\section{Introduction}

In classical data transmission, repeaters are used to amplify the
data signals (bits) when they become weaker during their propagation.
In long-distance optical-fiber systems, for instance, repeaters
are used to compensate for the intensity losses caused by scattering
and absorption of light pulses propagating along the fibre.
In contrast to classical information, the above procedure
is impossible to realize when the transmitted data signals carry
bits of quantum information (qubits).
In an optical-fiber system, a qubit can be encoded into a
single photon which cannot be amplified or cloned without destroying
quantum coherence associated with this qubit \cite{nat299, pla92}.
Therefore, the photon has to propagate along the entire length
of the fiber which causes an exponentially decreasing
probability to detect this photon at the end of the channel.

To avoid the exponential decay of a photon wave-packet
and preserve its quantum coherence,
the concept of a quantum repeater was proposed \cite{prl81}.
According to this concept, a large set of entangled photon pairs
is distributed over sufficiently short fiber segments.
The two protocols (i) entanglement purification \cite{prl76, prl77} 
and (ii) entanglement swapping \cite{swap} are employed to extend 
the short-distance entangled photon pairs over the entire length 
of the channel. With the help of entanglement swapping, two 
entangled pairs of neighboring segments are combined into one 
entangled pair, gradually increasing the distance of shared 
entanglement. The entanglement purification enables one to 
distill high-fidelity entangled pairs from a larger set of
low-fidelity entangled pairs by means of local operations
performed in each of the repeater nodes and classical
communication between these nodes. The resulting entangled photon 
pair distributed between the end points of the photonic channel 
can then be used for quantum teleportation \cite{prl70} or 
quantum key distribution \cite{rmp74}.

Owing to the fragile nature of quantum correlations
and inevitable photon loss in the transmission channel,
it still poses a serious challenge for experimentalists
to outperform the direct transmission of photons along the
fiber. Up to now, only individual
building blocks of a quantum repeater have been experimentally
demonstrated including bipartite entanglement purification 
\cite{prl90, nat443}, entanglement swapping \cite{prl96, pra71}, 
and entanglement distribution between two neighboring nodes 
\cite{nat454, sc316}. 
Nevertheless, motivated both by the impressive experimental
progress and theoretical advances, various revised and improved 
implementations of repeaters or its building-blocks have been 
proposed \cite{pra79, pra81, pra81a, prl105, pra82, prl104a,
pra81b, rmp83, pra84, lpr, pra83, prl96a, pra78b, prl101}.

\begin{figure}
\begin{center}
\includegraphics[width=0.48\textwidth]{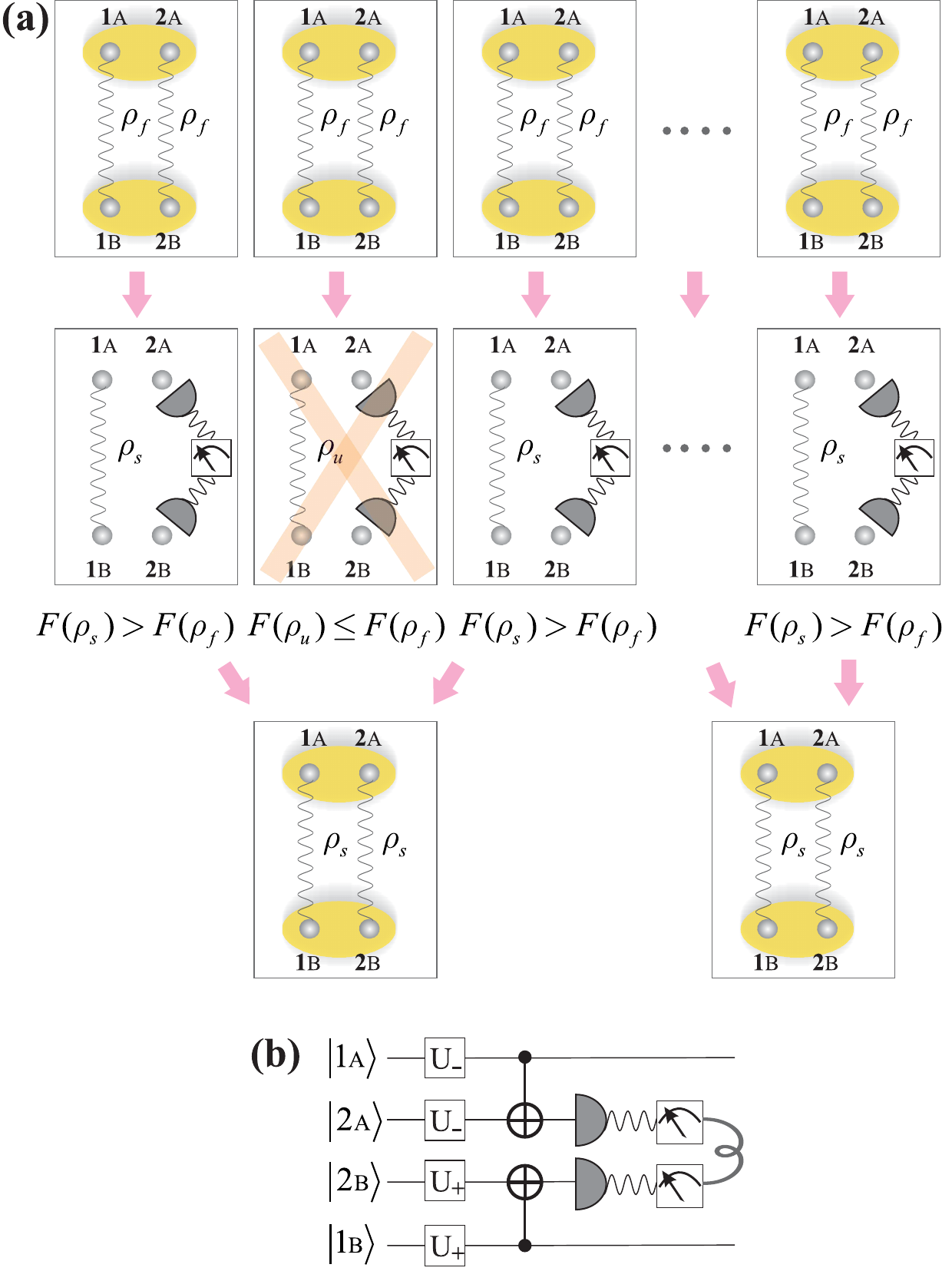} \\
\caption{(Color online) (a) Sequence of steps for the
conventional purification protocol described
in the text. In the upper part, repeater
nodes A and B share a set of low-fidelity entangled pairs
grouped into elementary blocks of four qubits, and the
grey ellipses indicate local interactions occurring
within each repeater node. In the middle part,
one qubit pair (inside each elementary block) is projectively
measured and the results are compared with a predefined outcome.
In the lower part, all successfully purified qubit pairs
are collected and the purification takes place one more time.
(b) Quantum circuit that corresponds to the interaction
indicated above by grey ellipses. See description of gates
in the text.}
\label{fig1}
\end{center}
\end{figure}

Practical schemes for implementing a quantum
repeater are not straightforward. The two mentioned protocols,
entanglement purification and entanglement swapping,
require feasible and reliable quantum logic,
such as single- and two-qubit gates.
Due to its complexity and high demand of physical resources,
entanglement purification is the most
delicate and cumbersome part of a quantum repeater.

In one of the purification protocols \cite{prl77}, two repeater
nodes A and B share a finite set of low-fidelity entangled pairs
grouped into elementary blocks of two qubit pairs as displayed
in Fig.~\ref{fig1}(a). Each entangled pair is given by the Werner
state \cite{pra40}
\begin{equation}\label{w-state}
\rho_f^{AB} = f \, \Phi^+_{AB}
            + \frac{1 - f}{3} \left( \Phi^-_{AB} + \Psi^+_{AB}
            + \Psi^-_{AB} \right),
\end{equation}
being diagonal in the usual Bell basis, with
$\Phi^\pm_{AB} \equiv \ket{\phi^\pm_{AB}} \bra{\phi^\pm_{AB}}$,
$\Psi^\pm_{AB} \equiv \ket{\psi^\pm_{AB}} \bra{\psi^\pm_{AB}}$.
The fidelity
\begin{equation}\label{f-def}
\mathbf{F}(\rho_f^{AB}) \equiv \text{Tr}\left[ \Phi^+_{AB} \, \rho_f^{AB} \right]
= f > 0.5
\end{equation}
is assumed to be above the threshold value of $1/2$. Both qubit
pairs (from each elementary block) are assumed to interact locally,
i.e., such that the interaction occurs only within one single repeater
node (A and B) as indicated by grey ellipses in Fig.~\ref{fig1}(a). 
More specifically, the qubit pairs $1_A - 2_A$ and $1_B - 2_B$ are first 
rotated with help of single-qubit gates $U_\pm = \frac{1}{\sqrt{2}}(I 
\pm \im \sigma_x)$, with the usual Pauli operators, and afterwards interact 
by means of CNOT gates as shown in Fig.~\ref{fig1}(b). 
By means of the latter gate, qubit $1_A$ 
($1_B$) acts as control qubit and qubit $2_A$ ($2_B$) as target qubit. 
While the control qubit does not change its state under the
CNOT gate, the target qubit is flipped once the control qubit is set
to the excited state. In order to simplify our further discussions, the
entire sequence of quantum gates indicated by grey ellipses
shall be referred to below as the purification gate. After the
purification gate is performed, qubits $2_A$ and $2_B$ are projected
on the computational basis $\{ \ket{0}, \ket{1} \}$ and the outcome
of these projections is exchanged between the two nodes by means of
classical communication [see the middle part of Fig.~\ref{fig1}(a)].

Entanglement purification is successful if the outcome of projections 
reads $\{ 0, 0 \}$ or $\{ 1, 1 \}$ for qubits $2_A$ and $2_B$. 
In this case, the (unmeasured) qubit pair $1_A - 1_B$ is described by the
Bell-diagonal density operator $\rho_s$ that implies the fidelity 
(we drop super- and subscripts AB)
\begin{equation}\label{f-cnot}
\mathbf{F}(\rho_s) \equiv \text{Tr}\left[ \Phi^+ \, \rho_s \right] =
\frac{1 - 2 \, f + 10 \, f^2}{5 - 4 \, f + 8 \, f^2} \, ,
\end{equation}
such that $\mathbf{F}(\rho_s) > \mathbf{F}(\rho_f)$ for any $f > 0.5$.
The entanglement purification is unsuccessful if the mentioned
outcome of projections reads $\{ 0, 1 \}$ or $\{ 1, 0 \}$ and the
qubit pair $1_A - 1_B$ must be discarded. Successfully purified pairs, 
in contrast, are collected from all blocks in order to carry out the 
next purification round as illustrated in the lower part of Fig.~\ref{fig1}(a).
At every purification round, high-fidelity entangled pairs are distilled 
from a larger set of low-fidelity pairs and the described procedure can 
be straightforwardly extended to more subsequent purification rounds once 
a sufficient amount of elementary blocks is provided at the input.

\begin{figure}
\begin{center}
\includegraphics[width=0.3\textwidth]{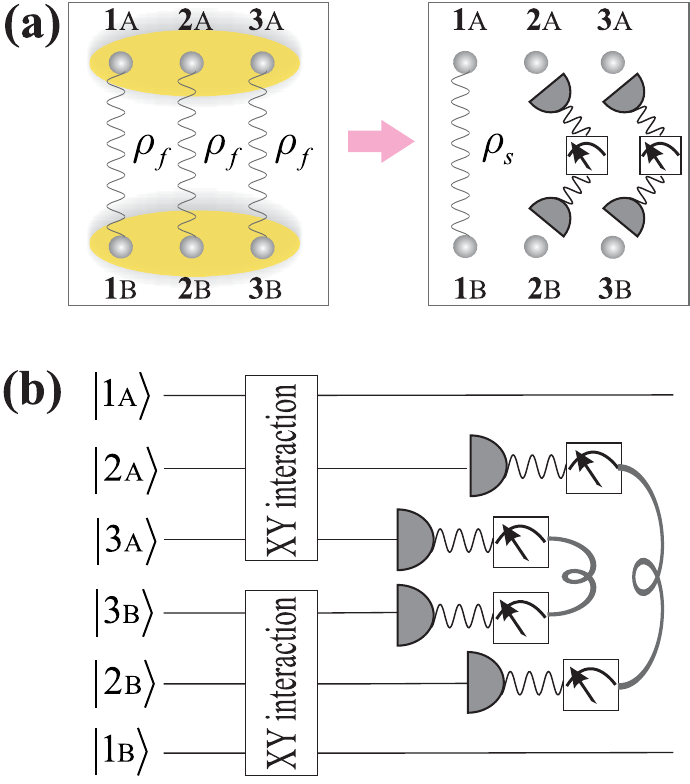} \\
\caption{(Color online) (a) Sequence of steps for the proposed
purification scheme using a single elementary block containing 
three low-fidelity entangled pairs. See text for description. 
(b) Quantum circuit that corresponds to the interaction depicted 
above by grey ellipses.}
\label{fig2}
\end{center}
\end{figure}

Obviously, any practical purification scheme has to be
resource-efficient and involve experimentally feasible qubit
operations. The above purification protocol, however,
involves cumbersome CNOT gates which pose a serious challenge
for most physical realizations of qubits, involving
complicated pulse sequences along with superfluous qubit operations
\cite{nat443, prl104, pra71a, prl85, pra78, pra67}.
In the current work, we present a more practical bipartite purification
scheme that exploits the {\it natural dynamics} of spin chains and
can be straightforwardly realized in the framework of cavity QED.
In contrast to conventional purification protocols (such as described 
above), in our purification scheme, (i) each elementary block contains 
three qubit pairs as displayed in Fig.~\ref{fig2}(a), (ii) single-qubit 
$U_\pm$ rotations together with CNOT gates are replaced by Heisenberg 
XY interactions as shown in Fig.~\ref{fig2}(b), and (iii) two qubit
pairs from each elementary block are projectively measured out after 
the interaction [see the right part of Fig.~\ref{fig2}(a)].
Entanglement purification is successful if the combined outcome of 
the projections coincides with a predefined outcome that is determined by 
the Heisenberg XY model (see Section \ref{sec-HXY}).

\begin{figure}[!t]
\begin{center}
\includegraphics[width=0.48\textwidth]{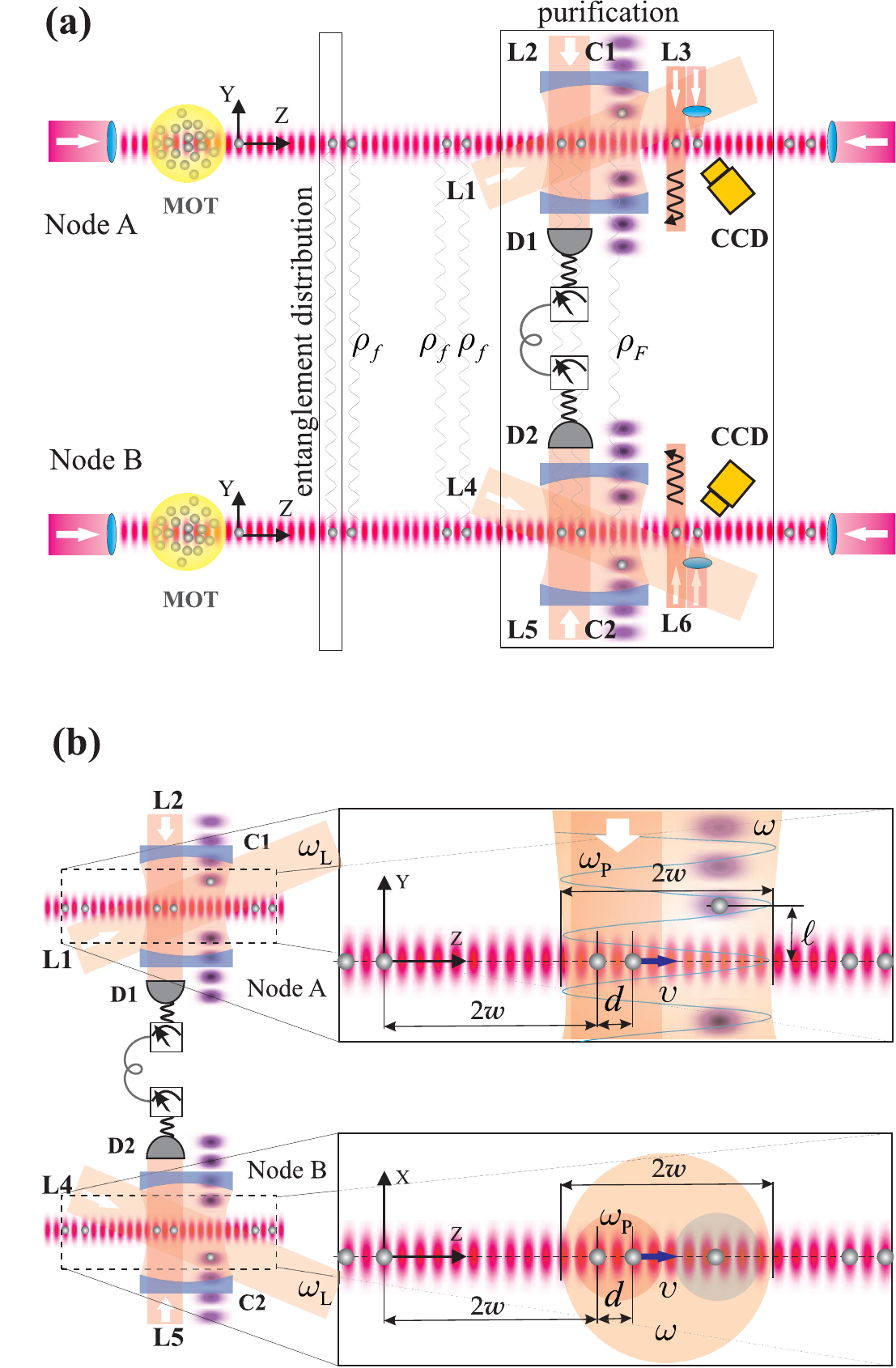} \\
\vspace{0.75cm}
\includegraphics[width=0.22\textwidth]{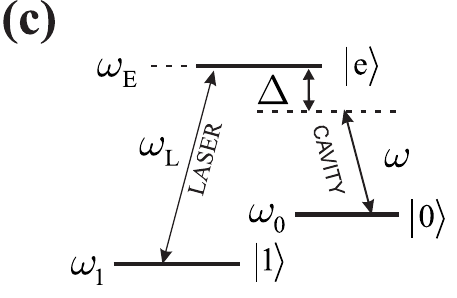} \\
\caption{(Color online) (a) Scheme of an experimental setup that
realizes the proposed purification scheme and is incorporated into 
a quantum repeater segment with two neighboring nodes. 
See text for description of components. (b) Detailed
view of the purification block that is indicated by a rectangle in
the above experimental setup. In the upper and lower parts of this figure
a magnification of side ($y-z$ plane) and top ($x-z$ plane) views are
shown, respectively. See text for description. (c) Structure of 
three-level atom in the $\Lambda$-type configuration.}
\label{fig3}
\end{center}
\end{figure}

Although we increase the number of qubit pairs inside each elementary 
block in our scheme, in every purification gate we avoid the direct use 
of CNOT gates together with single-qubit 
rotations and hence the need for complicated pulse sequences and extra 
qubit operations. The interaction based on the Heisenberg XY model,
moreover, describes the natural dynamics of spin chains that is 
characteristic for many physical realizations of qubits. It can be
deterministically obtained in the framework of cavity QED by using
short chains of atoms coupled to the same cavity mode of a high-finesse
optical resonator (see Refs.~\cite{prl85, prl87} for the simplest
case with two atoms).

In the present work, we develop an experimentally feasible purification 
scheme which exploits a cavity-mediated interaction between atoms that 
produces a Heisenberg XY type evolution. This results in a more resource- 
and time-efficient protocol if compared with conventional approaches.
Following the recent developments in 
cavity QED, moreover, we briefly point to and discuss a few practical 
issues related with the implementation of our purification scheme and
the main limitations which may arise on the experimental side. By 
combining our purification scheme with entanglement distribution and 
swapping protocols, a reasonably practical implementation 
of resource- and time-efficient quantum repeaters for long-distance 
quantum communication using chains of atoms and optical resonators
may be possible.

\begin{figure*}[!ht]
\begin{center}
\includegraphics[width=0.95\textwidth]{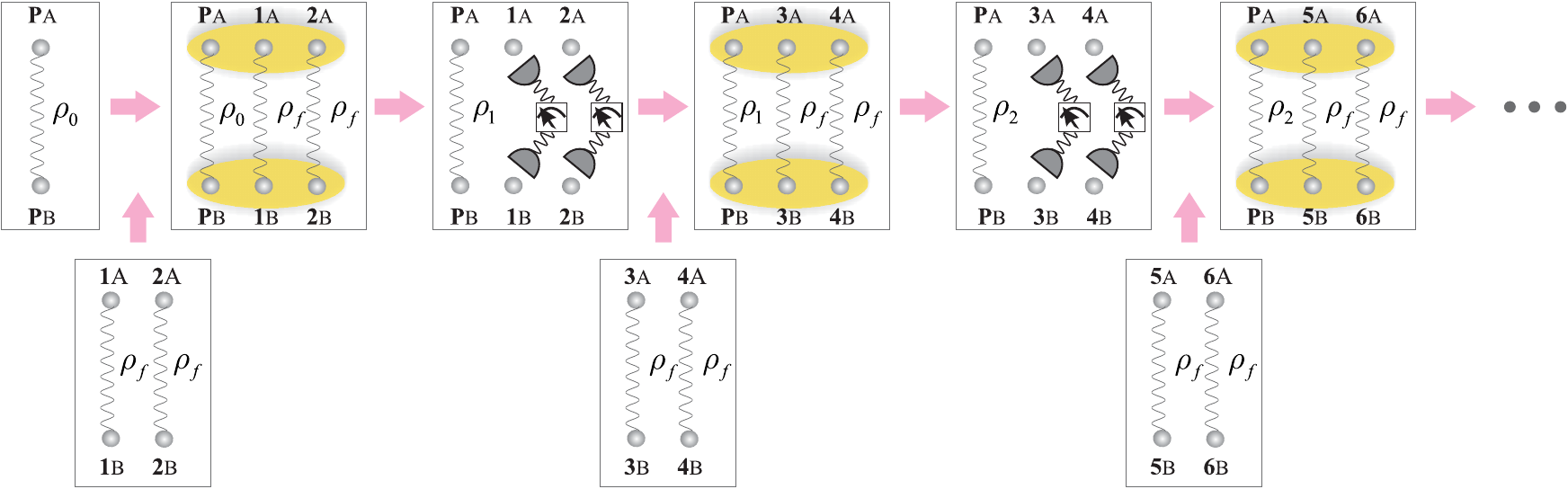} \\
\caption{(Color online) Sequence of steps for the modified
entanglement purification scheme that fits the experimental
setup from Fig.~\ref{fig3} and which utilizes one single
elementary block. This block contains three entangled qubit 
pairs: (i) one permanent qubit pair that encodes the distilled
qubits and (ii) two temporary qubit pairs which are employed 
to increase the entanglement fidelity of the permanent pair. 
See text for detailed description.}
\label{fig4}
\end{center}
\end{figure*}

The paper is organized as follows. In the next section, we 
first outline our purification scheme. We analyze the atomic evolution 
that is mediated by the cavity field in subsection II.A.  
In subsection II.B, we derive the effective Hamiltonian that governs 
this evolution and we identify it with the Heisenberg XY model. 
In subsection II.C, we apply the dynamics of the derived Hamiltonian 
to our purification scheme and determine the main properties which are 
relevant for our scheme. In subsections II.D - II.E, we discuss several
issues which follow from the analysis performed in the preceding 
subsections and which are crucial for our purification scheme. 
We discuss one issue related to the implementation of our purification 
scheme in subsection II.F, while a short summary and outlook are given
in section III.

\section{Purification Protocol without CNOT gates}

The main physical resources of our purification scheme are: 
(i) chains of atoms, (ii) high-finesse optical cavities,
and (iii) detectors for projective measurement of atomic states. 
In Fig.~\ref{fig3}(a) we show a scheme
of the proposed quantum repeater segment including two neighboring nodes 
(A and B) and combining entanglement purification and distribution protocols 
in a single experimental setup. In this setup, each repeater node consists 
of one optical cavity $C_1$ ($C_2$) acting along the $y$-axis, a laser beam
$L_1$ ($L_4$), a chain of atoms transported by means of an optical lattice 
along the along the $z$-axis, one stationary atom
trapped inside the cavity with the help of a vertical lattice,
laser beams $L_2$ ($L_5$) and $L_3$ ($L_6$) which act along the $y$-axis, 
a magneto-optical trap (MOT), a detector $D_1$ ($D_2$) connected to the 
neighboring node through a classical communication channel, and a CCD camera. 
First we shall connect the purification scheme introduced in the 
previous section with the experimental setup from Fig.~\ref{fig3}(a) to
clarify the role of each element.

Obviously, the sequence of steps in Fig.~\ref{fig1} with an elementary 
block displayed in Fig.~\ref{fig2} cannot be directly applied to
our experimental setup since it would require one
individual cavity for each elementary block and, therefore, an
unreasonable demand of physical resources. Instead, we shall
consider a modified sequence that is illustrated in Fig.~\ref{fig4} 
which perfectly fits into our proposed experimental setup. 
By this sequence, repeater nodes A and B share only one elementary 
block in which the qubit pair $P_A - P_B$ is a permanent pair and the 
other two pairs are introduced in the block temporarily in a successive
fashion. Initially, there is only one permanent pair in the block that
is characterized by the fidelity $\mathbf{F}(\rho_0) > 1/2$ and 
supplemented by two extra temporary pairs $1_A - 1_B$ and $2_A - 2_B$,
both characterized by the fidelity $\mathbf{F}(\rho_f) > 1/2$. 
Using these three qubit pairs, the sequence displayed in 
Fig.~\ref{fig2} can now be applied and in the case of successful 
purification, the increased fidelity $\mathbf{F}(\rho_1)$ 
of pair $P_A - P_B$, with $\mathbf{F}(\rho_1) > \mathbf{F}(\rho_0)$, is 
obtained. After the projected pairs $1_A - 1_B$ and $2_A - 2_B$ are 
replaced with two fresh entangled pairs $3_A - 3_B$ and $4_A - 4_B$, 
the purification steps are repeated.

While the permanent qubit pair encodes the distilled entangled qubit
pair and stores it for the subsequent purification rounds, temporary
qubit pairs are used to increase gradually the fidelity
of the permanent pair. Below we shall associate permanent qubits
with stationary atoms trapped inside cavities $C_1$ and $C_2$ and 
temporary qubits with atoms in the chain, inserted into the horizontal 
lattices and transported along the $z$-axis [see Fig.~\ref{fig3}(a)]. 
According to our experimental 
setup, this identification implies that atoms pass sequentially through 
the cavity and only two atoms from the chain can couple 
simultaneously to the same cavity mode. These two atoms together 
with the stationary (trapped) atom, therefore, provide an atomic triplet 
in each repeater node as required for our purification scheme.

Right before each atom from node A enters the cavity, it becomes
entangled with the respective atom from node B as depicted in
Fig.~\ref{fig3}(a) by wavy lines, such that each entangled
pair is described by Eqs.~(\ref{w-state})-(\ref{f-def}).
This entanglement is generated (non-locally) by means of an entanglement
distribution block, indicated in Fig.~\ref{fig3}(a) by a rectangle. 
This entanglement distribution may be realized in various ways 
\cite{prl78, nat453, prl96a}, where our purification protocol here does 
not depend on any specific choice for the entanglement distribution. 
During the transition of an atomic pair through the cavity, the triplet 
of atoms undergoes a cavity-mediated (Heisenberg XY) evolution 
in each of the repeater nodes, which shall be referred to below as the 
(new) purification gate. According to the sequence in Fig.~\ref{fig4}, 
furthermore, the purification sequence is completed once the states of 
a (conveyed) atomic pair are projectively measured and the outcome of 
projections is pairwise exchanged between the repeater nodes in order 
to decide if the purification was successful or not.

In our experimental scheme, the latter projections are performed by
means of a laser beam $L_3$ ($L_6$) and a CCD camera in each of the 
repeater nodes as displayed in Fig.~\ref{fig3}(a). While the laser 
beam $L_3$ ($L_6$) removes atoms in a given quantum state from the 
chain without affecting atoms in the other state (so-called push-out
technique \cite{prl91}), the CCD camera is used to detect the
presence of remaining atoms via fluorescence imaging and determine, 
therefore, the state of each atom that leaves the cavity. Assuming that 
the purification was successful, the next atomic pair from the chain 
enters the cavity (in each repeater node) and the next purification 
round occurs with the same stationary atom (permanent qubit). In the 
unsuccessful case, however, the stationary atoms should be re-initialized 
and the entire sequence from Fig.~\ref{fig4} should be re-started.

\subsection{Evolution of atoms due to cavity-mediated interaction}

We recall that each repeater node disposes $N$ atomic pairs
inserted into an optical lattice such that atoms within one pair are
separated by a distance $d$ and the distance between two neighboring
pairs is adjusted such that only one atomic pair can be
coupled simultaneously to the same cavity mode [see Fig.~\ref{fig3}(b)].
The entire chain is conveyed with a constant velocity $\upsilon$ along
the $z$-axis such that their position vectors
$\vec{r}_i$(t) = $\{0, 0, z_i^o + \upsilon \, t \}$ cross the cavity
at the anti-node ($y = 0$) and where $z_i^o$ denotes the initial position 
of the $i$-th atom outside the cavity. Both velocity $\upsilon$
and inter-atomic distance $d$ can be controlled experimentally by adjusting
the shift in the frequencies of the two counter-propagating laser beams
and by selecting a proper wavelength of the optical lattice, respectively
\cite{sc293}.

Apart from $N$ atomic pairs, moreover, each repeater node contains one
stationary atom trapped inside the cavity by a lattice acting along the
$y$-axis. As displayed in Fig.~\ref{fig3}(b), the position vector
$\vec{r}_s$ = $\{0, \ell, \ell \}$ of this atom crosses the cavity at the
anti-node ($y = \ell$) and is located in the same $y-z$ plane as the atomic 
chain. Each atom in our setup is a three-level atom in the $\Lambda$-type
configuration as displayed in Fig.~\ref{fig3}(c) encoding a single
qubit by means of states $\ket{0}$ and $\ket{1}$. In order to protect the 
qubit against decoherence caused by the fast-decaying excited state $\ket{e}$,
the states $\ket{0}$ and $\ket{1}$ are (typically) understood as stable 
ground and long-living metastable states or two hyperfine levels of the 
ground state.

We recall, furthermore, that the purification gate from Fig.~\ref{fig2}
is based on the Heisenberg XY interaction that is produced deterministically 
by coupling two-level atoms to the same mode of a high-finesse resonator
in our scheme. Due to the encoding scheme from Fig.~\ref{fig3}(c), however, 
the detuned optical cavity is coupled to the atomic transition
$\ket{0} \leftrightarrow \ket{e}$, while the atomic qubit is stored by
means of states $\ket{0}$ and $\ket{1}$. In order to couple the atomic qubit
to the cavity, therefore, an intermediate excitation coupled to the
$\ket{1} \leftrightarrow \ket{e}$ transition is further required in order 
to transfer coherently the electronic population of atoms from the 
qubit-storage states $\{ \ket{0}, \ket{1} \}$ to the cavity-active states 
$\{ \ket{0}, \ket{e} \}$. For this reason, the laser beam $L_1$ ($L_4$) 
is introduced in our experimental setup and coupled resonantly to the 
$\ket{1} \leftrightarrow \ket{e}$ transition.

Right after the atomic pair is loaded into the cavity, this laser
beam is switched on for a short time period that is equivalent to a
$\pi/2$ Rabi pulse. Under the action of this pulse, the laser field 
couples simultaneously to the atomic triplet (atomic pair from the 
chain and stationary atom) and transfers coherently the electronic 
population of each atom from the qubit-storage to cavity-active states. 
This coherent transfer, in turn, couples the atomic triplet to the cavity 
and activates the cavity-mediated evolution within this atomic triplet. 
In the same fashion, one additional laser pulse applied 
after the operational time that is required for the purification gate, 
transfers coherently the atomic population from the cavity-active to 
qubit-storage states backwards and stops the cavity-mediated evolution. 
By switching appropriately on and off the laser beam $L_1$ ($L_4$), 
therefore, we can precisely control the duration of the cavity-mediated 
evolution once the atomic pair is conveyed into the the cavity.

Before we turn to the Hamiltonian that governs the evolution 
of atoms coupled to the same cavity mode, it is important to explain 
the mechanism of cavity-mediated evolution of an atomic triplet. Let 
us consider, for instance, three atoms prepared initially in the
product state: $\ket{e_1, 0_2, 0_3}$, where the subscripts $1$ and
$2$ correspond to the atomic pair from the chain with position vectors
$\vec{r}_1(t)$ and $\vec{r}_2(t)$, respectively, while the subscript
$3$ corresponds to the stationary atom with position vector
$\vec{r}_3 \equiv \vec{r}_s$. In this case, composite atomic states
evolve according to the sequences:
\begin{equation}\label{seq1}
   \ket{e_1, 0_2, 0_3; \, \bar{0}} \rightarrow \ket{0_1, 0_2, 0_3; \, \bar{1}}
   \begin{array}{c}
     \nearrow \raisebox{0.15cm}{$\ket{0_1, e_2, 0_3; \, \bar{0}}$} \\
     \searrow \raisebox{-0.15cm}{$\ket{0_1, 0_2, e_3; \, \bar{0}}$}
   \end{array} \, ,
\end{equation}
if there were initially no photons in the cavity.

The middle part of the above sequence describes the cavity-mediated interaction
between the atoms realized by means of a single-photon exchange.
In order to avoid the (fast decaying)
cavity-excited state $\ket{0_1, 0_2, 0_3; \, \bar{1}}$ and ensure that the cavity
remains (almost) unpopulated during the entire evolution, we require a rather
large detuning between the atomic $\ket{0} \leftrightarrow \ket{e_i}$ transition
and the resonant frequency of the cavity field
\begin{equation}\label{cond}
|\,(\omega_E - \omega_0) - \omega \,| \gg g(\vec{r}_i), \quad i = 1,2,3
\quad,
\end{equation}
where $\omega$ is the resonant frequency of the cavity field and
\begin{equation}\label{coupling}
g(\vec{r}) = g_\circ \, \exp \left[ - (z^2 + x^2) / w^2 \right]
\end{equation}
is the (position-dependent) atom-cavity coupling\footnote{Gaussian
beam that is formed inside a cavity depends on the $y$-coordinate by
means of width of Gaussian beam, radius of curvature, and Guoy phase
\cite{gunther}. To a good approximation, however, we can neglect
this dependence if the distance between two cavity anti-nodes (to
which atomic pair and stationary atom are coupled) is relatively
small.}. This position-dependence is caused by the variation of the
transversal cavity field along atomic trajectories, where $g_\circ$
denotes the vacuum Rabi frequency and $w$ is the cavity field waist 
as seen in Fig.~\ref{fig3}(b).

With help of condition (\ref{cond}), the sequence (\ref{seq1}) reduces to
the effective sequence
\begin{equation}\label{seq2}
 \ket{e_1, 0_2, 0_3}
   \begin{array}{c}
     \nearrow \raisebox{0.15cm}{$\ket{0_1, e_2, 0_3}$} \\
     \searrow \raisebox{-0.15cm}{$\ket{0_1, 0_2, e_3}$}
   \end{array} \, ,
\end{equation}
where the (fast decaying) cavity-excited state has been omitted. 
By composing effective sequences for the remaining (three-qubit)
product states, we conclude that this evolution preserves the number
of excitations in the system and all composite atomic states can be
divided in four decoupled (non-overlapping) sub-spaces: (i)
$\ket{0_1, 0_2, 0_3}$, (ii) $\ket{e_1, 0_2, 0_3}$, $\ket{0_1, e_2,
0_3}$, $\ket{0_1, 0_2, e_3}$, (iii) $\ket{e_1, e_2, 0_3}$,
$\ket{0_1, e_2, e_3}$, $\ket{e_1, 0_2, e_3}$, and (iv) $\ket{e_1,
e_2, e_3}$. This conclusion implies, moreover, that the states from
groups (i) and (iv) remain trapped with regard to the cavity-mediated
evolution.

\subsection{Effective Hamiltonian associated with the cavity-mediated evolution}

While the effective sequence (\ref{seq2}) displays the
cavity-mediated evolution of three atoms, we still have to analyze
this evolution quantitatively in order to understand how to control
it in practice. For this purpose, we shall adiabatically eliminate
the intermediate state which we omitted in the effective sequence
and shall derive an effective Hamiltonian that governs the 
evolution associated with this sequence. For three identical atoms
which are coupled simultaneously to the same cavity mode, the
evolution of the combined triplet-cavity system is governed by 
the Jaynes-Cummings Hamiltonia
\begin{equation}\label{ham0}
H_{JC}(t) = - \hbar \, \Delta \, a^\dag \, a - \im \hbar \sum_{k=1}^3 \, g_k(t) \,
         \left( a \, \sigma_k^\dag - a^\dag \, \sigma_k \right).
\end{equation}
In this Hamiltonian, $a$  and $a^\dag$ denote the annihilation and
creation operators for a cavity photon that acts upon the
Fock states $\ket{\bar{n}}$, while $\sigma_i = \ket{0}_i \bra{e}$
and $\sigma_i^\dag =  \ket{e}_i \bra{0}$ are the atomic
lowering and raising operators, respectively. Moreover,
$\Delta = (\omega_E - \omega_0) - \omega$ refers to the detuning as
displayed in Fig.~\ref{fig3}(c) and the atom-cavity couplings
\begin{eqnarray}\label{couplings}
&& g_i(t) \equiv  g(\vec{r}_i) =
          g_\circ \, \exp \left[ - (z_i^o + \upsilon \, t)^2 / w^2 \right] ,
          \quad i = 1,2 \quad , \notag \\
&& \hspace{1.5cm}
g_3(t) \equiv  g(\vec{r}_s) = g_\circ \, \exp \left[ - \ell^2 / w^2 \right] ,
\end{eqnarray}
have been introduced, such that $|z_1^o - z_2^o| = d$.
The time evolution governed by this Hamiltonian, furthermore,
is described by the Schr\"{o}dinger equation
\begin{equation}\label{schrod_eq}
\im \hbar \, \frac{d \, \ket{\psi(t)}}{dt} = H_{JC}(t) \, \ket{\psi(t)}.
\end{equation}

As we explained above, all composite atomic
states are divided in four decoupled sub-spaces such that evolution
(\ref{schrod_eq}) preserves the number of excitations in the system.
Therefore, we can consider one particular ansatz for the wave-function
\begin{eqnarray}\label{ansatz}
\ket{\psi(t)} &=& c_0(t) \, \ket{0_1, 0_2, 0_3; \, \bar{1}}
               +  c_1(t) \, \ket{e_1, 0_2, 0_3; \, \bar{0}} \notag \\
              &+& c_2(t) \, \ket{0_1, e_2, 0_3; \, \bar{0}}
               +  c_3(t) \, \ket{0_1, 0_2, e_3; \, \bar{0}}, \quad 
\end{eqnarray}
that is based on sequence (\ref{seq1}). In the above ansatz,
we assume that $c_0(t), \ldots, c_{3}(t)$ are the complex and normalized
amplitudes, such that $c_0(0) = 0$. Using ansatz (\ref{ansatz}), 
the Schr\"{o}dinger equation (\ref{schrod_eq}) gives rise to
the closed system of equations
\begin{subequations}\label{solutions}
\begin{eqnarray}
   \im \dot{c}_0(t) &=& - \Delta \, c_0(t)
             + \im \sum_{k=1}^3 g_{k}(t) \, c_{k}(t), \label{s1} \\
   \dot{c}_i(t) &=& - g_{i}(t) \, c_{0}(t), \quad i = 1,2,3 \quad ,
   \label{s2}
\end{eqnarray}
\end{subequations}
which describes the evolution of the coupled atoms-cavity system that is 
governed by the Hamiltonian (\ref{ham0}), and where dot denotes the time
derivative.

According to the sequence (\ref{seq2}), the state $\ket{0_1, 0_2, 0_3; \, \bar{1}}$,
remains (almost) unpopulated once the atom-cavity detuning is chosen properly 
[see Eq.~(\ref{cond})]. In order to separate the evolution of unpopulated states
from Eqs.~(\ref{solutions}), we apply the adiabatic elimination procedure which 
assumes an adiabatic behavior of the amplitudes $c_0(t)$ and, hence, to a good 
approximation vanishing of its time derivatives (see, for instance, 
Ref.~\cite{jpa40}). We exploit this derivative $\dot{c}_0(t) \cong 0$ and obtain 
with help of Eq.~(\ref{s1}) an equation for $c_0(t)$ which we insert into 
Eq.~(\ref{s2}). The remaining three effective equations,
\begin{equation}\label{solutions1}
\im \dot{c}_{k}(t) = \sum_{j=1}^3 \frac{g_k(t) \, g_j(t)}
{\Delta} \, c_j(t) \, ,
\end{equation}
describe only the amplitudes $c_1(t)$, $c_2(t)$, $c_3(t)$ which correspond to
the states $\ket{e_1, 0_2, 0_3}$, $\ket{0_1, e_2, 0_3}$, $\ket{0_1, 0_2, e_3}$,
respectively.

Above equations which we derived from Eqs.~(\ref{solutions}) by using the
adiabatic elimination procedure, in turn, can be derived directly from the
Schr\"{o}dinger equation
\begin{equation}\label{schrod_eq1}
\im \hbar \, \frac{d \ket{\phi(t)}}{dt} = H(t) \, \ket{\phi(t)} \, ,
\end{equation}
\begin{equation}
\ket{\phi(t)} = c_1(t) \ket{e_1, 0_2, 0_3}
              + c_2(t) \ket{0_1, e_2, 0_3}
              + c_3(t) \ket{0_1, 0_2, e_3}, \notag
\end{equation}
associated with the effective Hamiltonian
\begin{eqnarray}
H(t) &=& \hbar \sum_{k=1}^3 \frac{g_k(t)^2}{\Delta} \ket{e}_k \bra{e} + \widetilde{H}(t) \, , 
\label{ham2} \\
\widetilde{H}(t) &=& \hbar \sum_{{i,j = 1} \atop{(i \neq j)}}^3 \frac{g_i(t) \, g_j(t)}{\Delta}
 \left( \sigma_i^\dag \sigma_j + \sigma_j^\dag \sigma_i \right), \label{ham-dd}
\end{eqnarray}
describing the evolution due to the effective sequence (\ref{seq2}), 
and where $\widetilde{H}(t)$ is the dipole-dipole interaction that 
describes the cavity-mediated energy exchange between the atoms.  
As a summary, we have shown that the evolution of an atomic
triplet coupled to the detuned cavity field is reduced to the
evolution of atoms which interact via single-photon exchange in such
a manner that the cavity-excited state remains (almost) unpopulated.

The Hamiltonian (\ref{ham2}) is complicated to handle analytically since 
it contains time-dependent (atom-cavity) couplings (\ref{couplings}). In order 
to simplify our further analysis and approximate reasonably well the evolution 
given by the Hamiltonian (\ref{ham2}), we calculate first the mean values of 
atom-cavity couplings and consider these instead of the time-dependent couplings 
in the above Hamiltonian. In order to proceed, we assume that the atomic triplet 
interacts with the cavity field during the entire transition time, in which the 
atomic pair is conveyed through the (waist region $w$ of the) cavity field. We note,
moreover, that the first term in the Hamiltonian (\ref{ham2}) doesn't contribute 
to the cavity-mediated energy exchange between the atoms that is essential for 
our protocol. Therefore, we ignore this term while calculating the mean values 
of atom-cavity couplings and we integrate the Eq.~(\ref{schrod_eq1}) with $H(t)$ 
being replaced by $\widetilde{H}(t)$
\begin{equation}\label{schrod_eq2}
\im \hbar \, \frac{d \, \ket{\phi(t)}}{dt} = \widetilde{H}(t) \, \ket{\phi(t)}.
\end{equation}
To a good approximation, the commutator $[ \widetilde{H}(t_1), \widetilde{H}(t_2) ]$ vanishes for 
all $t_1$ and $t_2$ because of a rather large detuning $\Delta$ in the denominator 
of Eq.~(\ref{ham-dd}) as compared to the quadratic atom-cavity coupling in the 
nominator [see Eq.~(\ref{cond})]. In a high-finesse cavity, moreover, the 
Gaussian envelope (\ref{couplings}) describes quite well the strength of the 
atom-cavity coupling and, therefore, we can safely integrate Eq.~(\ref{schrod_eq2}) 
from $t \rightarrow - \infty$ to $t \rightarrow + \infty$ 
\begin{equation}\label{evol0}
U(\infty) = \exp \left[ - \frac{\im}{\hbar} \int_{-\infty}^{\infty} \widetilde{H}(t)
            \, dt \right]
          = \exp \left[ - \frac{\im}{\hbar} \, H_\infty \, t^\prime \, \right],
\end{equation}
where $t^\prime \equiv \sqrt{\pi} \, w / \upsilon$ is the effective interaction 
time, and the asymptotic Hamiltonian $H_\infty$ is defined in the form
\begin{equation}\label{ham3}
H_\infty = \frac{\hbar \, g_\circ^2 \, e^{- \ell^2 / w^2}}{\Delta} \sum_{{i,j = 1}
      \atop{(i \neq j)}}^3 C_{ij}
      \left( \sigma_i^\dag \sigma_j + \sigma_j^\dag \sigma_i \right) \, ,
\end{equation}
with the coupling-terms $C_{13} = C_{31} = C_{23} = C_{32} = 1$,
\begin{equation}
\text{and} \qquad
C_{12} = C_{21} =
\frac{1}{\sqrt{2}} \exp \left[ \frac{2 \, \ell^2 - d^2}{2 \, w^2} \right].
\end{equation}

By inserting the mean atom-cavity couplings from the asymptotic Hamiltonian
(\ref{ham3}) into the effective Hamiltonian (\ref{ham2}), we obtain the mean
Hamiltonian in the form
\begin{equation}\label{ham4}
H_M =
  \frac{\hbar \,g^2}{\Delta} \left[ \sum_{k=1}^3 \ket{e}_k \bra{e} +
 \sum_{{i,j = 1} \atop{(i \neq j)}}^3
 \left( \sigma_i^\dag \sigma_j + \sigma_j^\dag \sigma_i \right) \right],
\end{equation}
where $g \equiv g_\circ \, e^{- \ell^2 /2 \, w^2}$ is the mean atom-cavity
coupling, and where we assumed that distances $d$ and $\ell$ are adjusted
such that $C_{12} = C_{21} = 1$. In an appropriate interaction picture,
furthermore, the mean Hamiltonian (\ref{ham4}) can be expressed in the
simplified form
\begin{eqnarray}\label{ham5}
H_I &=&
     \frac{\hbar \, g^2}{\Delta} \sum_{{i,j = 1} \atop{(i \neq j)}}^3
     \left( \sigma_i^\dag \sigma_j + \sigma_j^\dag \sigma_i \right) \notag \\
    &=&
     \frac{\hbar \, J}{2} \sum_{i = 1}^3
     \left( \sigma_i^x \sigma_{i+1}^x + \sigma_i^y \sigma_{i+1}^y \right),
\end{eqnarray}
which coincides with the isotropic Heisenberg XY interaction Hamiltonian
with periodic boundary conditions, i.e., $\sigma_4^x = \sigma_1^x$ and
$\sigma_4^y = \sigma_1^y$ \cite{ap16}, and where $J \equiv g^2 / \Delta$
is the coupling between three spins. This coupling, moreover, can be positive
or negative depending on the sign of the atom-cavity detuning $\Delta$
[see Fig.~\ref{fig3}(c)].

To summarize this section, we calculated first the
effective Hamiltonian (\ref{ham2}) starting from Eqs.~(\ref{schrod_eq}),
(\ref{ansatz}) with help of the adiabatic elimination procedure. Secondly,
we obtained the mean values of atom-cavity couplings from the asymptotic
Hamiltonian (\ref{ham3}) and derived the mean Hamiltonian (\ref{ham5}).
We stress, moreover, that the ansatz (\ref{ansatz}) is based on those composite 
atomic states from group (ii) which involve one single excitation. 
It can be shown, however, that Hamiltonian (\ref{ham5}) preserves the number 
of excitations in the system and it describes correctly the atomic evolution 
that is based on the composite atomic states from groups (i), (iii), 
and (iv).

\begin{figure}[!ht]
\begin{center}
\includegraphics[width=0.425\textwidth]{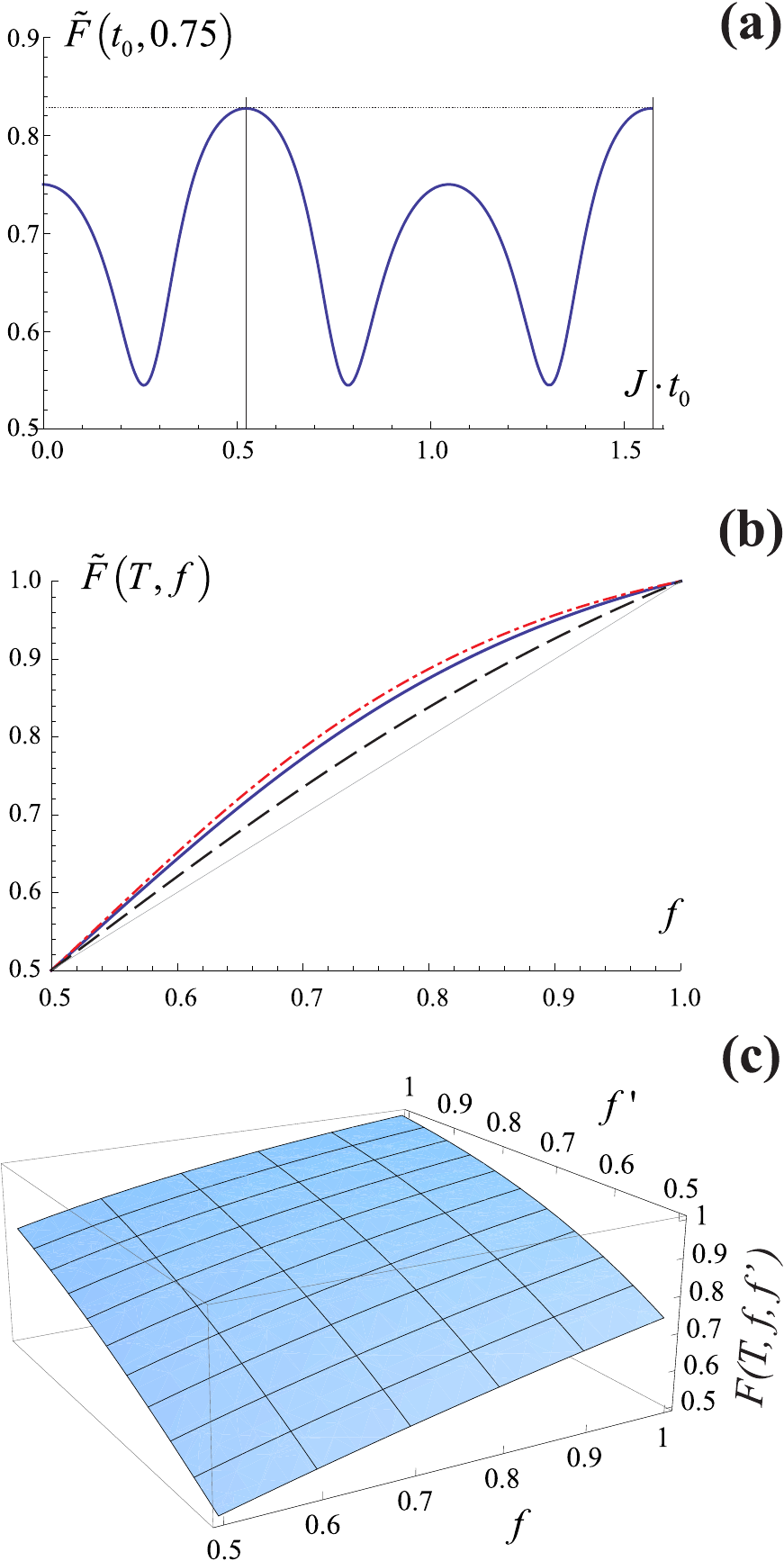} \\
\vspace{0.75cm}
\caption{(Color online) (a) Fidelity $\widetilde{F}(t, 0.75)$
as a function of operational time $t_0$ with fixed input fidelity $f = 0.75$.
(b) Fidelity $\widetilde{F}(T, f)$ (solid curve) as a function 
of input fidelity $f$ with fixed operational time given by
Eq.~(\ref{time}). Behavior of Eq.~(\ref{f-cnot}) (dashed curve) 
obtained in the conventional (CNOT-based) purification protocol after
one purification round. 
Fidelity (dot-dashed curve) obtained after two (successful) 
purification steps using the scheme C from Ref.~\cite{pra59} 
(see discussions in the text).
(c) Fidelity $F(T, f, f^\prime)$ as a function of input 
fidelities $f$ and $f^\prime$ associated with the conveyed atomic 
pairs and the stationary atomic pair, respectively, with the 
operational time $T$ given by Eq.~(\ref{time}).}
\label{fig5}
\end{center}
\end{figure}
\subsection{Dynamics of Heisenberg XY model and purification protocol}\label{sec-HXY}

In the previous subsection, we identified the cavity-mediated evolution
of an atomic triplet with the evolution of three spins
which interact by means of the isotropic Heisenberg XY Hamiltonian with
periodic boundary conditions. In the framework of cavity QED,
therefore, the purification gate that is required for our scheme can
be deterministically realized by coupling an atomic triplet to the
same (detuned) cavity mode for a predefined time period. In this
section, we shall analyze the evolution governed by the Heisenberg XY
Hamiltonian (\ref{ham5}) and determine its properties along with 
necessary operational times.

Earlier we explained that right before each atomic pair
from node A enters the cavity, it becomes entangled pairwise with another
atomic pair from node B as depicted in Fig.~\ref{fig3}(a) by wavy lines.
We denote two conveyed pairs of atoms in the nodes A and B with labels $1,2$
and $4,5$, respectively, while the stationary atoms are labeled by $3$ and $6$,
respectively. By this notation, the conveyed atomic pairs
are described by the density operators $\rho_{f}^{1,4}$ and $\rho_{f}^{2,5}$,
while the pair of stationary atoms is described by the density operator
$\rho_{f^\prime}^{3,6}$. Since each purification round should lead to 
a gradual growth of entanglement fidelity (for the stationary atoms), we 
distinguish the initial fidelities $f$ of conveyed atomic pairs from the 
fidelity $f^\prime$ of stationary atoms.

The evolution associated with the Hamiltonian (\ref{ham5}),
\begin{equation}\label{evol1}
U_I(t) = e^{- \frac{\im}{\hbar} H_I \, t} = \sum_{k=1}^8 e^{- \frac{\im}{\hbar} E_k \, t} \, \ket{k} \bra{k} \, ,
\end{equation}
is completely determined by the energies $E_k$ and vectors $\ket{k}$ given in 
the cavity-active basis $\{ \ket{0}, \ket{e} \}$, which satisfy the equality 
$H_I \, \ket{k} = E_k \, \ket{k}$ with orthogonality and completeness relations
$\braket{k}{k^\prime} = \delta_{k k^\prime}$ and $\sum \ket{k} \bra{k} = I$,
respectively. This eigenvalue problem has been exactly solved with help of 
the Jordan-Wigner transformation \cite{zp47} and its respective solutions 
have been presented in the literature (see, for instance, Ref.~\cite{pra64}).
Since the evolution operator (\ref{evol1}) acts on the states of one
atomic triplet that is pairwise entangled with another atomic triplet 
(in the neighboring node), we have to consider the composite
evolution operator
\begin{equation}\label{evol2}
U(t) = \sum_{k,k^\prime =1}^8 e^{- \frac{\im}{\hbar} \left( E_k + E_{k^\prime} \right) \, t}
          \, \ket{k_A \otimes k_B^\prime} \bra{k_A \otimes k_B^\prime}.
\end{equation}
According to this evolution operator, the state of both atomic triplets in 
nodes A and B is described by the six-qubit density operator
\begin{equation}\label{density0}
\rho^{1-6}(t, f, f^\prime) = U(t) \left( \rho_{f}^{1,4} \otimes \rho_{f}^{2,5}
                      \otimes \rho_{f^\prime}^{3,6} \right) U^\dag(t),
\end{equation}
where the states of each atomic triplet have been coherently mapped from
the qubit-storage basis $\{ \ket{0}, \ket{1} \}$ to the cavity-active basis 
$\{ \ket{0}, \ket{e} \}$ by means of the laser beam $L_1$ ($L_4$).

Recall that in order to finalize one purification round,
we map back the states of both atomic triplets from the cavity-active 
basis $\{ \ket{0}, \ket{e} \}$ to the qubit-storage basis 
$\{ \ket{0}, \ket{1} \}$ and project both (conveyed) atomic pairs 
once they leave their cavities. This implies that after an 
evolution time $t = t_0$, during which the atoms evolved due to
cavity-mediated interaction and that is followed by the mapping
$\{ \ket{0}, \ket{e} \} \rightarrow \{ \ket{0}, \ket{1} \}$,
the state of both atomic triplets in nodes A and B is described by 
the density operator
\begin{equation}\label{density3}
\tilde{\rho}^{1-6}(t_0, f, f^\prime) = 
     \sum_{i,j=1}^{64} \, \rho^{1-6}_{ij}(t_0, f, f^\prime) \, \ketv{i} \brav{j},
\end{equation}
where the composite six-qubit vectors $\ketv{i}$ are given in the
computational (qubit-storage) basis $\{ \ket{0}, \ket{1} \}$ and satisfy the 
orthogonality and completeness relations $\braketv{i}{j} = \delta_{ij}$ and 
$\sum \ketv{i} \brav{i} = I$, respectively. Once both (conveyed) atomic pairs
leave their cavities, the projective measurements of their states is performed 
and the whole sequence of steps leads to the density operator and the probability 
of success,
\begin{eqnarray}\label{density1}
&& \rho^{3,6}(t_0, f, f^\prime) = \frac{
      \sum_{\alpha, \beta} \, 
      \rho^{1-6}_{\alpha \beta}(t_0, f, f^\prime)}
      {P_{\text{succ}}(t_0, f, f^\prime)} \, \ketvt{\alpha} \bravt{\beta}, \\
&& P_{\text{succ}}(t_0, f, f^\prime) = 
      \text{Tr} \left[ \sum_{\alpha, \beta} 
      \rho^{1-6}_{\alpha \beta}(t_0, f, f^\prime)  \, 
      \ketvt{\alpha} \bravt{\beta} \right], \notag
\end{eqnarray}
that describes only the state of stationary atoms, and 
where the Greek indices run over four values $i_1,i_2,i_3,i_4$ or
$j_1,j_2,j_3,j_4$ such that
$\ketvt{\alpha} \equiv \langle 0_1, 1_2, 0_4, 1_5 | \mathbf{V}_\alpha \rangle \neq 0$
or $\ketvt{\alpha} \equiv \langle 1_1, 0_2, 1_4, 0_5 | \mathbf{V}_\alpha \rangle \neq 0$,
respectively. As we shall see below, one of the two outcomes (of projective
measurements),
\begin{equation}\label{outcome}
\{ 0_1, 1_2, 0_4, 1_5 \} \quad \text{or} \quad \{ 1_1, 0_2, 1_4, 0_5 \} \, ,
\end{equation}
is the necessary condition to complete successfully one purification
round and increase the entanglement fidelity associated with 
the stationary atoms.

According to the six-qubit density operator (\ref{density0}), we have
routinely computed the matrix elements $\rho^{1-6}_{\alpha \beta}(t_0, f, f^\prime)$
which, however, are rather bulky to be displayed here. The first
relevant result we obtain with help of these matrix elements is
that the density operator (\ref{density1}) preserves the diagonal
form in the Bell basis,
\begin{equation}\label{density2}
\rho^{3,6}(t_0, f, f^\prime) = F \, \Phi^+_{36} +
          \frac{1 - F}{3} \left( \Phi^-_{36} + \Psi^+_{36} + \Psi^-_{36} \right).
\end{equation}
Unlike the conventional purification protocol, therefore, the
purified state (\ref{density2}) is a Werner state that is completely
characterized by the fidelity $\mathbf{F} \left( \rho^{3,6} \right)
= F(t_0, f, f^\prime)$, where the functions $\widetilde{F}(t_0,
f) \equiv F(t_0, f, f)$ and $\widetilde{P}_{\text{succ}}(t_0, f) \equiv 
P_{\text{succ}}(t_0, f, f)$ have the form
\begin{widetext}
\begin{eqnarray}
&& \hspace{1cm} \widetilde{F}(t_0, f) = \frac{f - 38 f^2 -8 + 8 (1 - 5 f + 4 f^2)
\cos(6 \, J \, t_0) - 12 f (4 f - 1) \cos(12 \, J \, t_0)}{34 f - 
32 f^2 - 47 + 16 (1 - 5 f + 4 f^2) \cos(6 \, J \, t_0) - 4 (2 f
+ 8 f^2 - 1) \cos(12 \, J \, t_0)} \, , \label{function} \\
&& \widetilde{P}_{\text{succ}}(t_0, f) =  (1 + 2f) \left[ 47 - 34
f + 32 f^2 - 16 (1 - 5 f + 4 f^2) \cos(6 \, J \, t_0) + 4 (2 f
+ 8 f^2 - 1) \cos(12 \, J \, t_0) \right] / 972 \, . \quad
\end{eqnarray}
\end{widetext}

In Fig.~\ref{fig5}(a), we display $\widetilde{F}(t_0, f)$ as a
function of $t_0$ for the input fidelity $f = 0.75$. It is 
clearly seen that after the time interval $t_0 = T$ that satisfies 
$(n = 0, 1, 2, \ldots)$,
\begin{equation}\label{time}
J \, T = \frac{\pi}{3} \left( n + \frac{1}{2} \right) \, ,
\end{equation}
this fidelity reaches its maximum value $\cong 0.827$. This
equality, therefore, sets the operational time that is
required for execution of the purification gate and which makes
the output fidelity $\widetilde{F}(T, f)$ increase with regard to 
its input value $f$. Corresponding to this operational time, furthermore, 
the fidelities $F(T, f, f^\prime)$ and $\widetilde{F}(T,f)$ along with 
the success probabilities $P_{\text{succ}}(T, f, f^\prime)$ 
and $\widetilde{P}_{\text{succ}}(T, f)$ take the following simplified 
form,
\begin{eqnarray}
&& F(T, f, f^\prime) = \frac{f^\prime (12 \,
f + 236 \, f^2 - 5) - 16 (f - 1)}{59 + (12 - 64 \, f^\prime) f - 4
(5 - 64 f^\prime) f^2},  \notag \\
&& \hspace{1.5cm} \widetilde{F}(T, f) = \frac{16 - 53 \, f + 118 \,
f^2}{59 - 106 \, f + 128 \, f^2} \, ,  \label{fidelity} \\
&& P_{\text{succ}}(T, f, f^\prime) =  
    \frac{59 + (12 - 64 f^\prime) f + 4 (-5 + 64 f^\prime) f^2}{972} 
    \, , \notag \\ 
&& \quad \widetilde{P}_{\text{succ}}(T, f) = 
    (59 + 12 \, f - 84 \, f^2 + 256 \, f^3) / 972 \, . 
    \quad \label{success}
\end{eqnarray}

The first two expressions describe quantitatively how the input
fidelity of stationary atoms is modified due to one single (and
successful) round of our purification scheme. In
Fig.~\ref{fig5}(b), we compare the fidelity $\widetilde{F}(T, f)$ 
(solid curve) with the fidelity given by Eq.~(\ref{f-cnot}) (dashed 
curve) that is obtained due to a single successful purification 
round in the conventional (CNOT-based) protocol. 
This comparison shows that the increase of
fidelity in our scheme is almost twice as large as for the
conventional protocol. This nice result, however, originates merely 
from the fact that our scheme requires one extra qubit pair for a 
single purification round. Being projectively measured right after the 
first pair, this extra pair leads to a stronger entanglement distillation
of the stationary atoms in the case of successful purification.

In order to motivate the latter conclusion, recall that we considered 
the purification sequence displayed in Fig.~\ref{fig4} that fits 
perfectly our experimental setup from Fig.~\ref{fig3}. In fact, this 
sequence is similar to scheme C of W.~D\"{u}r and co-authors, as
presented in Ref.~\cite{pra59}, using CNOT gates and freshly prepared
entangled pairs for every single purification step (so-called 
{\it entanglement pumping}). 
In this (CNOT-based) scheme, therefore, the number of 
atomic pairs required for two purification rounds (one permanent 
pair and two successive temporary pairs) coincides with the number of 
atomic pairs as required for one single purifications round in our 
scheme. For this equal amount of atomic resources, we plotted in 
Fig.~\ref{fig5}(b) the fidelity (dot-dashed curve) obtained after 
two (successful) purification rounds using the scheme C by 
W.~D\"{u}r and co-authors. It is clearly seen that this fidelity 
deviates only slightly from the fidelity (solid curve) obtained
after one single purification round in our scheme.

Apparently, our purification scheme is useful 
only if each round leads to a gradual growth of entanglement 
fidelity (of yjr stationary atoms) with regard to the respective 
fidelity obtained in the previous round, i.e.,
\begin{equation}\label{seq3}
f^\prime < F_{1} (T, f, f^\prime) < F_{2} (T, f, F_{1}) < \ldots < F_{n}
(T, f, F_{n-1}). 
\end{equation}
In Fig.~\ref{fig5}(c), we displayed the output fidelity $F(T, f, f^\prime)$  
obtained for one single and successful purification round. 
This fidelity exhibits growth for (one and the same) input fidelity $f$ 
of the conveyed atomic pairs and (growing) fidelity 
$f^\prime = F_{n-1}(T, f, F_{n-2})$ of the stationary 
atomic pair, which was obtained in the previous purification round. 
The behavior of output fidelity $F_{n}(T, f, F_{n-1})$ that is obtained 
in the $n$-th purification round, therefore, is in agreement 
with the sequence (\ref{seq3}) and ensures that each 
successful round leads to a gradual growth of fidelity.

\subsection{Remarks on the entanglement distribution between stationary atomic qubits}

Recall that right before each atomic pair from node A enters the cavity, 
it becomes entangled (pairwise) with another atomic pair from node B, such
that the conveyed atomic pairs are described by the density operators 
$\rho_{f}^{1,4}$ and $\rho_{f}^{2,5}$. We assumed, moreover, that the pair 
of stationary atoms is initially described by the density operator 
$\rho_{f^\prime}^{3,6}$, however, without explaining how this entangled 
state is initially created.

In this section, we suggest that there is no need to introduce an additional
entanglement distribution device in our experimental setup in order to 
entangle stationary atoms prior to the purification rounds. Instead, 
we prepare initially two atoms in the product state 
$\rho^{3,6}_0 = \ket{0_3, 0_6} \bra{0_3, 0_6}$ and start our purification 
protocol. It can be shown that the first successful round 
transforms the above product state into an entangled state described by
\begin{eqnarray}\label{state}
\rho^{3,6}_f &=& A(f) \, \Phi^+_{3,6} 
               + B(f) \, \left( \Psi^+_{3,6} + \Psi^-_{3,6} \right) 
               + C(f) \, \Phi^-_{3,6} \notag \\
             &+& D(f) \, \left( \ket{\phi^+_{3,6}} \bra{\phi^-_{3,6}} + 
                             \ket{\phi^-_{3,6}} \bra{\phi^+_{3,6}} \right)
\end{eqnarray}
being off-diagonal in the Bell basis with $A(f) \cong f > 1/2$, such that
$\mathbf{F}(\rho^{3,6}_f) > \mathbf{F}(\rho^{3,6}_0)$.

Obviously, the above state is no longer a Werner state like in 
Eq.~(\ref{density2}). We verified, however, that the function $D(f)$, which 
describes off-diagonal contributions, becomes negligibly small after a few 
successful rounds. Therefore, we succeed to entangle remotely two 
stationary atoms by means of local (in each repeater node) interactions, 
such that the input fidelity $f$ of the conveyed atomic pairs is mapped 
(almost) completely to the output fidelity of the stationary atomic pair. 
The output fidelity $A(f) \cong f$, plays the role of the input 
fidelity $f^\prime$ for the next purification round and leads to a gradual 
growth of entanglement fidelity in agreement with the sequence (\ref{seq3}).
We have verified, moreover, that the output fidelities due to the initial state 
(\ref{state}) coincide (almost) with the output fidelities $F_n(T,f,F_{n-1})$ 
due to the state (\ref{density2}). The price we pay for one extra (successful) 
purification round prior to the main sequence of rounds, therefore, 
is efficiently compensated by a more moderate demand of physical resources 
in our purification scheme.

\begin{figure}[t]
\begin{center}
\includegraphics[width=0.455\textwidth]{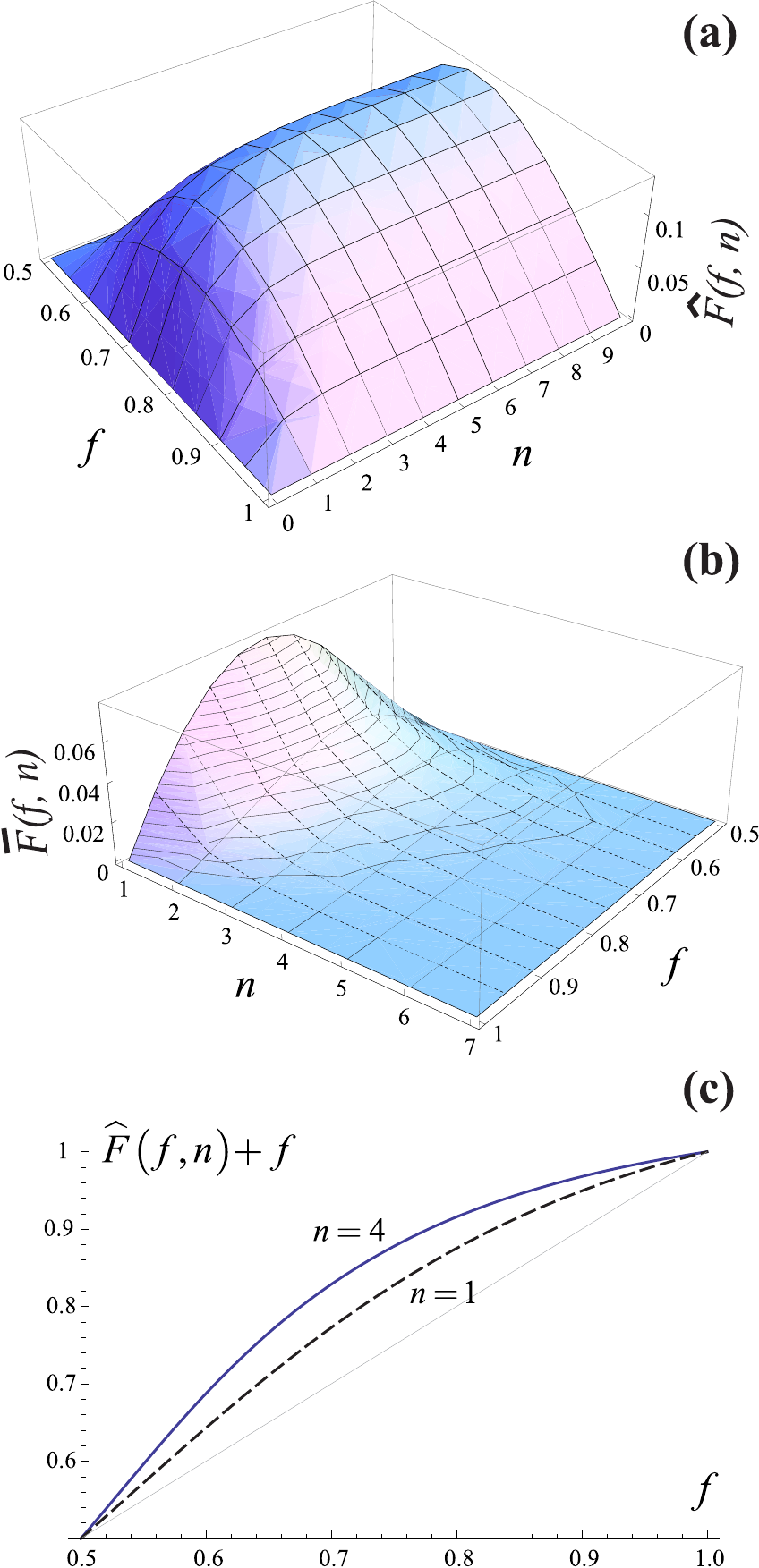} \\
\vspace{0.75cm}
\caption{(Color online) Behavior of (a) $\widehat{F}(f, n)$ and 
(b) $\bar{F}(f,n)$ as functions of input fidelity $f$ and number $n$ 
of purification rounds. (c) Final fidelity $\widehat{F}(f, n) + f$ 
as a function of initial fidelity $f$ with fixed numbers: $n=4$ 
(solid curve) and $n=1$ (dashed curve) of purification rounds.}
\label{fig6}
\end{center}
\end{figure}
\subsection{Saturation of entanglement purification and optimal number of rounds}

Recall that instead of the sequence from Fig.~\ref{fig1},
we considered a modified sequence displayed in Fig.~\ref{fig4}, 
similar to the scheme C that was proposed and discussed by W.~D\"{u}r and 
co-authors with help of CNOT gates \cite{pra59}. Using this scheme,
it has been pointed out that regardless of the number of (successful) 
purification rounds, the final fidelity is bounded by a fixed point 
that is smaller than the respective point due to the sequence from 
Fig.~\ref{fig1}. Similarly, the final fidelity $F_n (T, f, F_{n-1})$ 
that is obtained at the end of sequence (\ref{seq3}) is also bounded by 
a fixed point in our scheme. In this section, we shall refer to this
property as the saturation of entanglement purification and we will calculate
the optimal number of (successful) purification rounds that is required 
to reach closely this fixed point in a resource-efficient way.

While the sequence (\ref{seq3}) displays the gradual growth of entanglement
fidelity, we still have to analyze this growth quantitatively in order to 
understand how much the output fidelity increases with each purification 
round. For this purpose, we shall consider the following sequence,
\begin{equation}\label{seq4}
f < F_{1}(T, f, f) < \ldots < F_{n}(T, f, F_{n-1}) \equiv f + \widehat{F}(f, n) \, .
\end{equation}
In Fig.~\ref{fig6}(a), we show the function 
$\widehat{F}(f, n)$ that describes the difference between the final fidelity 
$F_{n}(T, f, F_{n-1})$ after $n$ (successful) rounds and the initial 
fidelity $f$ ($n = 0$). It is clearly seen that during the first four successful
rounds, this function exhibits a notably fast growth which, however, 
saturates afterwards and yields a rather minor 
increase with regard to the fidelity $F_{4}(T, f, F_{3})$.

In order to estimate the relative growth of fidelity that is obtained with
each purification round, in Fig.~\ref{fig6}(b) we display the 
function 
\begin{equation}\label{def}
\bar{F}(f, n) = F_{n}(T, f, F_{n-1}) - F_{n-1}(T, f, F_{n-2}) \, ,
\end{equation}
where $F_0 \equiv f$. This function shows quantitatively how much the 
output fidelity increases with each purification round in dependence 
upon both the input fidelity $f$ and the number of rounds $n$. This time it 
is clearly seen that the relative growth of fidelity has its maximum for 
$f = 0.75$ and it mostly vanishes after six successive rounds for all values 
of $f$. In dependence upon the input fidelity $f$, therefore, the surface 
from Fig.~\ref{fig6}(b) enables one to determine the optimal number of 
(successful) purification rounds $n_f$ that is required to reach closely 
the fixed point given by $\widehat{F}(f, n_f) + f$.

Corresponding to $n=4$ purification rounds, for which the final fidelity 
reaches its saturation level, in Fig.~\ref{fig6}(c), we display the 
fidelity $\widehat{F}(f, n) + f$ for $n=4$ rounds (solid curve) and the
fidelity obtained for $n=1$ round (dashed curve).
By comparing these two curves, we conclude that the increase of 
final fidelity due to four successive rounds is notably larger if compared 
to the case of one single purification round. We stress, finally, that W.~D\"{u}r 
and co-authors have demonstrated in Ref.~\cite{pra59} that 
errors and faulty quantum operations may lead to a situation, in which the 
fixed point in scheme C oversteps the respective fixed point of the conventional 
protocol based on the sequence from Fig.~\ref{fig1}. We expect that 
a similar behavior may exist in our scheme as well, however, this analysis 
is beyond the scope of the present paper.

\subsection{Remarks on the implementation of our purification scheme}\label{sec-RPP}

In our purification scheme, short chains of two atoms (in each 
repeater node) have to be transported with a constant velocity along the 
experimental setup and coupled to the cavity field in a well controllable 
fashion. For this purpose, two basic devices are required to store the atoms 
and transport them coherently 
into the cavity, namely, (i) a magneto-optical trap (MOT) that plays the
role of an atomic source and (ii) an optical lattice (conveyor belt) that 
transports atoms into the cavity from the MOT with a certain position and 
velocity control over the atomic motion. These two tools, combined 
with a high-finesse optical cavity in the same experimental setup 
\cite{prl95, prl98, njp10}, enable one to store initiated atoms in the MOT 
and insert them into the optical lattice for further transportation 
through the cavity. It has been experimentally demonstrated that an optical 
lattice preserves the coherence of transported atoms and can be utilized as a 
holder of a quantum register. Moreover, by encoding the quantum information 
by means of hyperfine atomic levels, a storage time of the order of seconds 
has been reported in Refs.~\cite{prl91, prl93}. The number-locked insertion 
technique \cite{njp12}, furthermore, allows to extract atoms from the MOT and 
insert a predefined pattern of them into an optical lattice with a single-site 
precision.

\section{Summary and Discussion}

In this paper, an experimentally feasible scheme was proposed to purify 
dynamically entanglement of two atoms which are trapped in two remote 
optical cavities. Our scheme utilizes chains of low-fidelity 
entangled atomic pairs which are coupled sequentially to both remote optical 
cavities. In contrast to conventional purification protocols, 
we avoid CNOT gates and hence reduce the need for complicated pulse sequences 
and superfluous qubit operations. Our purification mechanism exploits: 
(i) cavity-mediated interactions between atoms producing
Heisenberg XY evolutions governed by the Hamiltonian (\ref{ham5}) and 
(ii) projective measurements of atomic states. 
A detailed experimental setup was proposed in Fig.~\ref{fig3} and a complete 
description of all necessary steps and manipulations was provided. 
A comprehensive analysis of the fidelity obtained after multiple 
purification rounds was performed and the optimal number of rounds 
was determined by means of Fig.~\ref{fig6}. Following recent developments 
in cavity QED, moreover, we briefly pointed to and discussed a few practical 
issues related to the implementation of our purification scheme, including the 
main limitations which may arise on the experimental side. We stress that 
although the proposed purification scheme is experimentally feasible, its 
complete realization still poses a serious challenge.

Being the most delicate and cumbersome part of a quantum repeater, entanglement 
purification as proposed in this paper opens a route towards 
practical implementations of resource- and time-efficient quantum repeaters 
for long-distance quantum communication using chains of atoms and 
optical resonators. In our 
experimental setup, each atom from node A is assumed to become entangled 
with another atom from node B right before they enter their respective cavities. 
This entanglement is generated (non-locally) with 
help of an entanglement distribution block as indicated in Fig.~\ref{fig3}(a) 
by a rectangle. Since we consider only atoms and optical resonators as the 
physical resources for our repeater protocol, we suggest for entanglement 
distribution the scheme proposed in Ref.~\cite{prl96a} that allows
to entangle non-locally two (three-level) atoms in neighboring 
repeater nodes.

By this scheme, a coherent-state light pulse interacts with the first 
coupled atom-cavity system in node A, such that the optical field accumulates 
a phase conditioned upon the atomic state. After this, the 
light pulse propagates to repeater node B, where it interacts with the 
second coupled atom-cavity system and accumulates another phase conditioned 
upon the state of the atom located in this node. Finally, the phase-rotated coherent 
state is measured via homodyne detection and an entangled state is non-locally
generated through postselection between the atoms belonging to repeater nodes. 
The controlled phase rotation required for this scheme can be realized by 
means of a dispersive interaction of a single atom coupled to a high-finesse cavity 
in each repeater node. A detailed inclusion of this entanglement distribution 
scheme into our experimental setup is beyond the scope of this paper and will be  
subject of our next work.

In a recent paper by K.~Maruyama and F.~Nori \cite{pra78a}, a purification 
mechanism similar to ours and based on the natural dynamics of spin chains has been 
proposed. In contrast 
to our approach, however, the purification mechanism of that reference exploits the 
Heisenberg XYZ model with open boundary conditions which cannot be straightforwardly 
realized in the framework of cavity QED. 
Although the time behavior of our output fidelity $\widetilde{F}(t, f)$ is different 
from the respective time behavior found by K.~Maruyama and F.~Nori (compare 
Fig.~\ref{fig5}(a) with Fig.~3 from Ref.~\cite{pra78a}), the Heisenberg XYZ dynamics 
leads to the same expression (\ref{fidelity}) for a properly chosen interaction 
time. Apart from the simplified Hamiltonian utilized in our work, 
perfectly adapted to the cavity QED framework, the operational time that 
is required for one purification gate in our scheme is three times faster  
compared to the Heisenberg XYZ-based dynamics employed by K.~Maruyama and F.~Nori.

Finally, we would like to mention Ref.~\cite{qip} by A.~Casaccino and co-authors
in which an entanglement purification protocol based on the Heisenberg XY interaction 
has also been considered. In contrast to our scheme and the scheme of K.~Maruyama 
and F.~Nori, however, in that reference an abstract approach based on the spin 
dynamics on networks of various topologies has been analyzed. Using this general 
approach, the authors concluded that the maximum efficiency of entanglement 
purification is obtained only by means of networks with no isolated nodes. The 
requirement of no isolated nodes corresponds to the periodic boundary conditions 
in our approach [see Eq.~(\ref{ham5})] and, therefore, we confirmed in our paper 
the main standpoints of A.~Casaccino and co-authors from a particular point of view 
using cavity QED settings.

\begin{acknowledgments}

We thank the DFG for support through the Emmy Noether program. In addition,
we thank the BMBF for support through the QuOReP program. 

\end{acknowledgments}

\end{document}